\documentclass[aps,prl,twocolumn,showpacs,superscriptaddress,preprintnumbers,floatfix]{revtex4-1}  
\usepackage{graphicx}  
\usepackage{dcolumn}   
\usepackage{bm}        
\usepackage{amssymb}   
\usepackage{xspace}
\usepackage{multirow}
\usepackage{color}
\usepackage{units}

\newcommand{\lenopot}{\ensuremath{10.56\times10^{20}}\xspace}

\newcommand{\baseline}{\unit[735]{km}}
\newcommand{\ndbaseline}{\unit[1.04]{km}}

\newcommand{\dmsq}[1]{\ensuremath{\Delta m^2_{ #1 }}\xspace}

\newcommand{\numu}{\ensuremath{\nu_{\mu}}\xspace}          
\newcommand{\numubar}{\ensuremath{\overline{\nu}_{\mu}}\xspace}   
 
\newcommand{\nus}{\ensuremath{\nu_{s}}\xspace}          

\newcommand{\nue}{\ensuremath{\nu_{e}}\xspace}                      
\newcommand{\nuebar}{\ensuremath{\overline{\nu}_{e}}\xspace}        
\newcommand{\nutau}{\ensuremath{\nu_{\tau}}\xspace}              

\begin{document}

\pacs{14.60.Pq, 14.60.Lm, 29.27.-a}

\title{Search for Sterile Neutrinos Mixing with Muon Neutrinos in MINOS}

\newcommand{\Berkeley}{Lawrence Berkeley National Laboratory, Berkeley, California, 94720 USA}
\newcommand{\Cambridge}{Cavendish Laboratory, University of Cambridge, 
Cambridge CB3 0HE, United Kingdom}
\newcommand{\Cincinnati}{Department of Physics, University of Cincinnati, Cincinnati, Ohio 45221, USA}
\newcommand{\FNAL}{Fermi National Accelerator Laboratory, Batavia, Illinois 60510, USA}
\newcommand{\RAL}{Rutherford Appleton Laboratory, Science and Technology Facilities Council, Didcot, OX11 0QX, United Kingdom}
\newcommand{\UCL}{Department of Physics and Astronomy, University College London, 
London WC1E 6BT, United Kingdom}
\newcommand{\Caltech}{Lauritsen Laboratory, California Institute of Technology, Pasadena, California 91125, USA}
\newcommand{\Alabama}{Department of Physics and Astronomy, University of Alabama, Tuscaloosa, Alabama 35487, USA}
\newcommand{\ANL}{Argonne National Laboratory, Argonne, Illinois 60439, USA}
\newcommand{\Athens}{Department of Physics, University of Athens, GR-15771 Athens, Greece}
\newcommand{\NTUAthens}{Department of Physics, National Tech. University of Athens, GR-15780 Athens, Greece}
\newcommand{\Benedictine}{Physics Department, Benedictine University, Lisle, Illinois 60532, USA}
\newcommand{\BNL}{Brookhaven National Laboratory, Upton, New York 11973, USA}
\newcommand{\CdF}{APC -- Universit\'{e} Paris 7 Denis Diderot, 10, rue Alice Domon et L\'{e}onie Duquet, F-75205 Paris Cedex 13, France}
\newcommand{\Cleveland}{Cleveland Clinic, Cleveland, Ohio 44195, USA}
\newcommand{\Delhi}{Department of Physics \& Astrophysics, University of Delhi, Delhi 110007, India}
\newcommand{\GEHealth}{GE Healthcare, Florence South Carolina 29501, USA}
\newcommand{\Harvard}{Department of Physics, Harvard University, Cambridge, Massachusetts 02138, USA}
\newcommand{\HolyCross}{Holy Cross College, Notre Dame, Indiana 46556, USA}
\newcommand{\Houston}{Department of Physics, University of Houston, Houston, Texas 77204, USA}
\newcommand{\IIT}{Department of Physics, Illinois Institute of Technology, Chicago, Illinois 60616, USA}
\newcommand{\Iowa}{Department of Physics and Astronomy, Iowa State University, Ames, Iowa 50011 USA}
\newcommand{\Indiana}{Indiana University, Bloomington, Indiana 47405, USA}
\newcommand{\ITEP}{High Energy Experimental Physics Department, ITEP, B. Cheremushkinskaya, 25, 117218 Moscow, Russia}
\newcommand{\JMU}{Physics Department, James Madison University, Harrisonburg, Virginia 22807, USA}
\newcommand{\LASL}{Nuclear Nonproliferation Division, Threat Reduction Directorate, Los Alamos National Laboratory, Los Alamos, New Mexico 87545, USA}
\newcommand{\Lebedev}{Nuclear Physics Department, Lebedev Physical Institute, Leninsky Prospect 53, 119991 Moscow, Russia}
\newcommand{\Lancaster}{Lancaster University, Lancaster, LA1 4YB, UK}
\newcommand{\LLL}{Lawrence Livermore National Laboratory, Livermore, California 94550, USA}
\newcommand{\LosAlamos}{Los Alamos National Laboratory, Los Alamos, New Mexico 87545, USA}
\newcommand{\Manchester}{School of Physics and Astronomy, University of Manchester, 
Manchester M13 9PL, United Kingdom}
\newcommand{\MIT}{Lincoln Laboratory, Massachusetts Institute of Technology, Lexington, Massachusetts 02420, USA}
\newcommand{\Minnesota}{University of Minnesota, Minneapolis, Minnesota 55455, USA}
\newcommand{\Crookston}{Math, Science and Technology Department, University of Minnesota -- Crookston, Crookston, Minnesota 56716, USA}
\newcommand{\Duluth}{Department of Physics, University of Minnesota Duluth, Duluth, Minnesota 55812, USA}
\newcommand{\Ohio}{Center for Cosmology and Astro Particle Physics, Ohio State University, Columbus, Ohio 43210 USA}
\newcommand{\Otterbein}{Otterbein University, Westerville, Ohio 43081, USA}
\newcommand{\Oxford}{Subdepartment of Particle Physics, University of Oxford, Oxford OX1 3RH, United Kingdom}
\newcommand{\PennState}{Department of Physics, Pennsylvania State University, State College, Pennsylvania 16802, USA}
\newcommand{\PennU}{Department of Physics and Astronomy, University of Pennsylvania, Philadelphia, Pennsylvania 19104, USA}
\newcommand{\Pittsburgh}{Department of Physics and Astronomy, University of Pittsburgh, Pittsburgh, Pennsylvania 15260, USA}
\newcommand{\IHEP}{Institute for High Energy Physics, Protvino, Moscow Region RU-140284, Russia}
\newcommand{\Rochester}{Department of Physics and Astronomy, University of Rochester, New York 14627 USA}
\newcommand{\RoyalH}{Physics Department, Royal Holloway, University of London, Egham, Surrey, TW20 0EX, United Kingdom}
\newcommand{\Carolina}{Department of Physics and Astronomy, University of South Carolina, Columbia, South Carolina 29208, USA}
\newcommand{\SDakota}{South Dakota School of Mines and Technology, Rapid City, South Dakota 57701, USA}
\newcommand{\SLAC}{Stanford Linear Accelerator Center, Stanford, California 94309, USA}
\newcommand{\Stanford}{Department of Physics, Stanford University, Stanford, California 94305, USA}
\newcommand{\StJohnFisher}{Physics Department, St. John Fisher College, Rochester, New York 14618 USA}
\newcommand{\Sussex}{Department of Physics and Astronomy, University of Sussex, Falmer, Brighton BN1 9QH, United Kingdom}
\newcommand{\TexasAM}{Physics Department, Texas A\&M University, College Station, Texas 77843, USA}
\newcommand{\Texas}{Department of Physics, University of Texas at Austin, 
Austin, Texas 78712, USA}
\newcommand{\TechX}{Tech-X Corporation, Boulder, Colorado 80303, USA}
\newcommand{\Tufts}{Physics Department, Tufts University, Medford, Massachusetts 02155, USA}
\newcommand{\UNICAMP}{Universidade Estadual de Campinas, IFGW, CP 6165, 13083-970, Campinas, SP, Brazil}
\newcommand{\UFG}{Instituto de F\'{i}sica, 
Universidade Federal de Goi\'{a}s, 74690-900, Goi\^{a}nia, GO, Brazil}
\newcommand{\USP}{Instituto de F\'{i}sica, Universidade de S\~{a}o Paulo,  CP 66318, 05315-970, S\~{a}o Paulo, SP, Brazil}
\newcommand{\Warsaw}{Department of Physics, University of Warsaw, 
PL-02-093 Warsaw, Poland}
\newcommand{\Washington}{Physics Department, Western Washington University, Bellingham, Washington 98225, USA}
\newcommand{\WandM}{Department of Physics, College of William \& Mary, Williamsburg, Virginia 23187, USA}
\newcommand{\Wisconsin}{Physics Department, University of Wisconsin, Madison, Wisconsin 53706, USA}
\newcommand{\deceased}{Deceased.}

\affiliation{\ANL}
\affiliation{\Athens}
\affiliation{\BNL}
\affiliation{\Caltech}
\affiliation{\Cambridge}
\affiliation{\UNICAMP}
\affiliation{\Cincinnati}
\affiliation{\FNAL}
\affiliation{\UFG}
\affiliation{\Harvard}
\affiliation{\HolyCross}
\affiliation{\Houston}
\affiliation{\IIT}
\affiliation{\Indiana}
\affiliation{\Iowa}
\affiliation{\Lancaster}
\affiliation{\UCL}
\affiliation{\Manchester}
\affiliation{\Minnesota}
\affiliation{\Duluth}
\affiliation{\Otterbein}
\affiliation{\Oxford}
\affiliation{\Pittsburgh}
\affiliation{\RAL}
\affiliation{\USP}
\affiliation{\Carolina}
\affiliation{\Stanford}
\affiliation{\Sussex}
\affiliation{\TexasAM}
\affiliation{\Texas}
\affiliation{\Tufts}
\affiliation{\Warsaw}
\affiliation{\WandM}

\author{P.~Adamson}
\affiliation{\FNAL}


\author{I.~Anghel}
\affiliation{\Iowa}
\affiliation{\ANL}



\author{A.~Aurisano}
\affiliation{\Cincinnati}









\author{G.~Barr}
\affiliation{\Oxford}









\author{M.~Bishai}
\affiliation{\BNL}

\author{A.~Blake}
\affiliation{\Cambridge}
\affiliation{\Lancaster}


\author{G.~J.~Bock}
\affiliation{\FNAL}


\author{D.~Bogert}
\affiliation{\FNAL}




\author{S.~V.~Cao}
\affiliation{\Texas}

\author{T.~J.~Carroll}
\affiliation{\Texas}

% ...

\collaboration{The MINOS Collaboration}
\noaffiliation

\date{\today}
\preprint{arXiv:1607.01176 [hep-ex]}
\preprint{FERMILAB-PUB-16-233-ND}

\begin{abstract}

We report results of a search for oscillations involving a light sterile neutrino over distances of 1.04 and \baseline{} in a \numu-dominated beam with a peak energy of \unit[3]{GeV}. The data, from an exposure of \lenopot{} protons on target, are analyzed using a phenomenological model with one sterile neutrino. We constrain the mixing parameters $\theta_{24}$ and \dmsq{41} and set limits on parameters of the four-dimensional Pontecorvo-Maki-Nakagawa-Sakata matrix, $|U_{\mu 4}|^{2}$ and $|U_{\tau 4}|^{2}$, under the assumption that mixing between \nue and \nus is negligible ($|U_{e4}|^{2}=0$). No evidence for $\numu \to \nus$ transitions is found and we set a world-leading limit on $\theta_{24}$ for values of $\dmsq{41} \lesssim \unit[1]{eV^{2}}$.
\end{abstract}

\maketitle

Studies of neutrinos and antineutrinos produced in the Sun, the atmosphere, and by reactors and accelerators~\cite{ref:PDG} have established that neutrinos have mass and that the weak-interaction flavor eigenstates $\nu_\ell$ ($l=e,\mu,\tau$) are related to the mass eigenstates $\nu_i$ ($i=1,2,3$) by a mixing matrix $U$:
\begin{equation}
 | \nu_\ell \rangle = \sum_{i} U_{\ell i} | \nu_i \rangle.
 \end{equation}

Measurements of the shape of the $Z$-boson resonance~\cite{ALEPH:2005ab} show that there are three active neutrino flavors with masses less than $m_Z/2$. The standard picture of neutrino mixing therefore assumes $U$ is a $3\times 3$ matrix, the Pontecorvo-Maki-Nakagawa-Sakata (PMNS) matrix~\cite{ref:MNS,ref:Pontecorvo,ref:Gribov}, that relates the flavor states to three neutrino mass states $m_{1}$, $m_{2}$, and $m_{3}$. The matrix is commonly parametrized using three mixing angles, $\theta_{12}$, $\theta_{23}$, and $\theta_{13}$, and a charge-parity violating phase $\delta$~\cite{Harari:1986xf}. The three angles and the two mass splittings $\dmsq{21}=m_{2}^{2}-m_{1}^{2}$
and $|\dmsq{32}|=|m_{3}^{2}-m_{2}^{2}|$
have been measured in multiple experiments~\cite{ref:PDG}.

The three-flavor model of neutrino mixing provides an excellent description of most, but not all, neutrino data. In particular, the Liquid Scintillator Neutrino Detector (LSND) observed a $3.8\sigma$ excess consistent with $\numubar \to \nuebar$ oscillations driven by a mass splitting $0.2 \leq \dmsq{\ } \leq \unit[10]{eV^2}$ that is incompatible with \dmsq{21} or \dmsq{32}~\cite{ref:LSND}. The MiniBooNE experiment searched for oscillations in the same range of mass splittings using beams of $\numu$ and $\numubar$ and found $3.4\sigma$ and $2.8\sigma$ excesses of \nue and \nuebar, respectively~\cite{ref:MiniBooNE}.

Many experiments have measured \nuebar fluxes from reactors at short baselines of $10\textrm{--}\unit[1000]{m}$.
A recent calculation~\cite{Mueller:2011nm,ref:HuberReactorFlux} predicts a flux that is about $3\%$ larger than previously assumed. The data display a deficit with respect to that prediction, the ``reactor anomaly,'' which can be interpreted as $\nuebar$ disappearance due to oscillations with $\dmsq{\ } \gtrsim\unit[0.1]{eV^2}$~\cite{ref:ReactorAnomaly}. Finally, a deficit of \nue has been observed from the gallium calibration sources of SAGE and GALLEX~\cite{ref:GalliumAnomaly,ref:GalliumStatistics}, which, when interpreted as oscillations, is consistent with the \dmsq{\ } range favored by the reactor anomaly. 

The anomalous oscillation signals described above may potentially be reconciled with data supporting the three-flavor oscillation picture by the addition of one or more sterile neutrinos that do not experience the weak interaction, but which mix with the active neutrinos~\cite{ref:SterileNeutrinoGlobalFit}. Since neutrinos have mass, sterile states may naturally arise from extensions to the standard model~\cite{ref:SterileNeutrinosExpected}. In this Letter, we test a phenomenological model in which the PMNS matrix is extended by the addition of a fourth neutrino mass eigenstate $\nu_4$ and a single sterile flavor state $\nu_s$. This ``3+1'' phenomenological model introduces three new mixing angles $\theta_{14}$, $\theta_{24}$, and $\theta_{34}$, and two additional phases $\delta_{14}$ and $\delta_{24}$ when parameterized as in Ref.~\cite{Harari:1986xf}. In this nomenclature the PMNS phase $\delta \equiv \delta_{13}$ and all $\delta_{ij}$-dependent terms appear multiplied by the corresponding $\sin\theta_{ij}$ in $U$.  In the following discussion we denote individual elements of $U$ as $U_{l i}$ with $l=e,\mu,\tau,s$ and $i=1,\ldots,4$. We also write $c_{ij}=\cos\theta_{ij}$, $s_{ij}=\sin\theta_{ij}$, and $\Delta_{ji}=\frac{\dmsq{ji}L}{4E_\nu}$, where $\dmsq{ji}\equiv m^{2}_{j}-m^{2}_{i}$, $L$ is the distance traveled by the neutrino, and $E_{\nu}$ is the neutrino energy.

The MiniBooNE and LSND experiments were conducted at $L/E_\nu \sim \unit[1]{km/GeV}$, a parameter space in which $\sin^2\Delta_{32} \sim 10^{-5}$ and $\sin^2\Delta_{21} \sim 10^{-8}$, rendering oscillations due to $\dmsq{32}$ and $\dmsq{21}$ negligible. In this case, and assuming $|\dmsq{41}| \gg |\dmsq{32}| > |\dmsq{21}|$, the $\nue$ appearance probability is
\begin{equation} \label{eq:miniboone}
P(\numu \to \nue) = 4 |U_{\mu 4}|^2 |U_{e 4}|^2 \sin^2 \Delta_{41},
\end{equation}
where $|U_{\mu 4}| = c_{14}s_{24}$ and $|U_{e 4}| = s_{14} $. 
Reactor experiments study $\nuebar \to \nuebar$ and have placed stringent limits on $\theta_{14}$~\cite{ref:DayaBay8AD,ref:Bugey}.

\begin{figure}
\begin{center}
\includegraphics[width=\columnwidth]{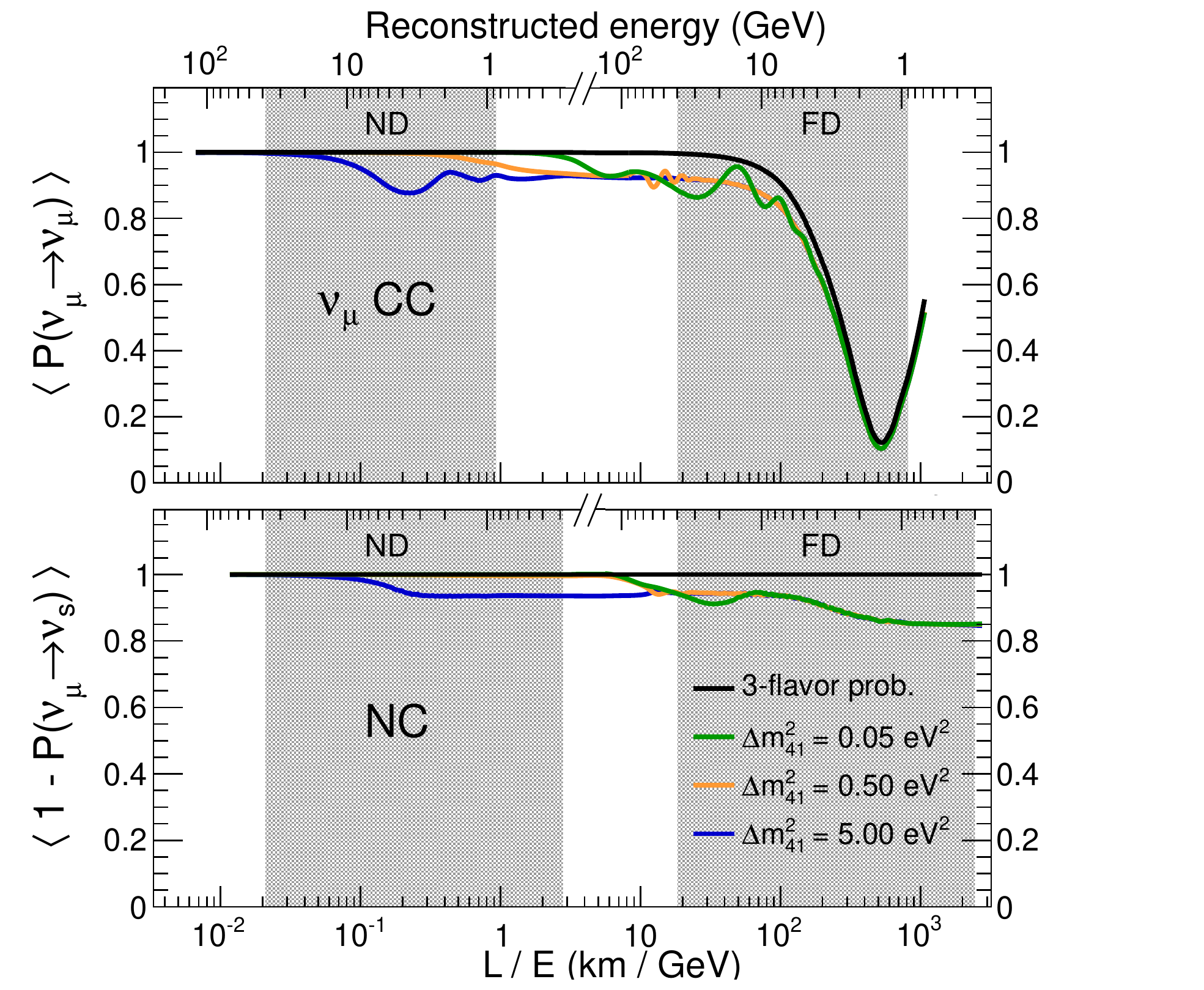}
\end{center}
\caption{\label{fig:osc} Muon neutrino oscillation probabilities as a function of $L/E$, where $L$ is the distance traveled by the neutrinos, and $E$ is the reconstructed neutrino energy (top horizontal axis of each panel), for three different values of $\dmsq{41}$, with $\theta_{14}=0.15$, $\theta_{24}=0.2$, $\theta_{34}=0.5$, and values of $\dmsq{31}$, $\dmsq{21}$, $\theta_{12}$, $\theta_{23}$, and $\theta_{13}$ from Ref.~\cite{ref:PDG}. The dip in $P(\numu \to \numu)$ at \unit[500]{km/GeV} is due to oscillations driven by \dmsq{31}. As $L/E$ increases, the various oscillation probabilities become similar and the lines overlap. The gray bands indicate the regions of reconstructed energy where CC \numu interactions (top panel) and NC interactions (bottom panel) are observed in the two detectors.
}
\end{figure}

MINOS measures neutrino oscillations using \numu charged-current (CC) and neutral-current (NC) interactions in a far detector (FD) and a near detector (ND) separated by \unit[734]{km}~\cite{ref:minosAppearanceAndDisappearance,ref:MINOSFinalDisappearance}. The neutrinos are produced by directing protons with energies of \unit[120]{GeV} from the Fermilab Main Injector onto a graphite target, located \ndbaseline{} upstream of the ND, producing $\pi$ and $K$ mesons. These mesons are focused by magnetic horns before decaying in a \unit[675]{m} long tunnel to produce predominantly muon-type neutrinos~\cite{ref:NuMIBeamPaper}. The ranges of $L/E$ probed by the two MINOS detectors are shown in Fig.\ \ref{fig:osc}. Disappearance of \numu occurs with a probability
\begin{equation} \label{eq:numu_osc}
P(\numu \to \numu) = 1-4\sum_{i=1}^4 \sum_{j>i}^4 |U_{\mu i}|^2 |U_{\mu j}|^2 \sin^2 \Delta_{ji}.
\end{equation}
In the analysis presented in this Letter, we use the exact oscillation probability to extract limits on the parameters. In the following discussion of the phenomenology, for simplicity we only show leading terms.

Terms in $\Delta_{21}$ are negligible, and we can approximate $\dmsq{32}\approx\dmsq{31}$. In the limit $\dmsq{41}\gg\dmsq{31}$ we can also approximate $\dmsq{43}\approx\dmsq{42}\approx\dmsq{41}$ and expand the oscillation probability to second order in the small terms $s_{13},s_{14},s_{24}$ and $\cos 2\theta_{23}$, yielding
\begin{eqnarray} \label{eq:cc}
P(\numu \to \numu)  \approx  1 & - & \sin^2 2\theta_{23} \cos 2\theta_{24}\sin^2\Delta_{31} \nonumber \\* 
& - &\sin^2 2\theta_{24} \sin^2\Delta_{41}.
\end{eqnarray} 
Thus, mixing with sterile neutrinos in the MINOS CC \numu sample is controlled by $\theta_{24}$ and would be seen as a depletion of events for $\Delta_{41}\gtrsim \pi/2$, as shown in the top panel of Fig.~\ref{fig:osc}.

For $10^{-3}\lesssim\dmsq{41}\lesssim\unit[0.1]{eV^2}$ an energy-dependent depletion would be observed at the FD with no effect at the ND. The $\Delta m^{2}_{41}=\unit[0.05]{eV^{2}}$ curve in the top panel of Fig.~\ref{fig:osc} shows an example of this behavior. As \dmsq{41} increases toward $\unit[1]{eV^2}$ we have $\Delta_{41}\gg \pi/2$ at the FD. In this case---the fast-oscillation regime---an energy-independent reduction in the event rate would be observed, since $\sin^{2}\Delta_{41} \to \nicefrac{1}{2}$ when the finite energy resolution of the detectors is considered. The $\Delta m_{41}^{2}=\unit[0.50]{eV^{2}}$ curve in the top panel of Fig.~\ref{fig:osc} shows an example of fast oscillations. For $\dmsq{41}\gtrsim\unit[1]{eV^2}$ an additional energy-dependent depletion of \numu would be seen at the ND, with the energy of maximum oscillation increasing with \dmsq{41}. An example of these ND oscillations is shown by the $\Delta m^{2}_{41}=\unit[5.00]{eV^{2}}$ curve in the top panel of Fig.~\ref{fig:osc}. For $\dmsq{41}\gtrsim\unit[100]{eV^2}$ fast oscillations occur at both detectors.

MINOS is also sensitive to sterile neutrinos via the disappearance of NC events~\cite{ref:FirstMINOSNCPRL,ref:MINOSNCPRD,ref:SecondMINOSNCPRL}, as shown in the bottom panel of \mbox{Fig.\ \ref{fig:osc}}, which would occur with a probability
\begin{eqnarray}\label{eq:nc}
1 - P(\numu \to \nus) & \approx & 1 - c_{14}^4 c_{34}^2 \sin^2 2\theta_{24} \sin^2 \Delta_{41} \nonumber \\*
& & - A \sin^2 \Delta_{31} + B \sin 2\Delta_{31}.
\end{eqnarray}
The terms $A$ and $B$ are functions of the mixing angles and phases. To first order, $A=s^2_{34}\sin^2 2\theta_{23}$ and $B=\frac{1}{2}\sin\delta_{24}s_{24} \sin 2\theta_{34} \sin 2\theta_{23}$. The NC sample is therefore sensitive to $\theta_{34}$ and $\delta_{24}$ in addition to $\theta_{24}$, although that sensitivity is limited by poor neutrino-energy resolution (due to the undetected outgoing neutrino), a lower event rate due to cross sections, and \numu and \nue CC backgrounds.

The MINOS apparatus and NuMI beam have been described in detail elsewhere~\cite{ref:minosnim,ref:NuMIBeamPaper}. We analyze an exposure of \lenopot protons on target (POT) used to produce a \numu-dominated beam with a peak energy of \unit[3]{GeV}.  The detectors are magnetized steel-scintillator, tracking-sampling calorimeters that utilize an average field of \unit[1.3]{T} to measure the charge and momentum of muons. The energy of hadronic showers is measured using calorimetry. In the case of CC \numu interactions, this is combined with topological information through a $k$-nearest-neighbor algorithm~\cite{ref:BackhouseThesis}.

A sample of NC-enhanced events is isolated by searching for interactions that induce activity spread over fewer than 47 steel-scintillator planes. Events with a reconstructed track are required to penetrate no more than five detector planes beyond the end of the hadronic shower. Additional selection requirements are imposed in the ND to remove cases in which the reconstruction program was confused by multiple coincident events. The selected NC sample in the ND has an efficiency of 79.9\% and a purity of 58.9\%, both estimated from Monte Carlo (MC) simulation. The background is composed of 86.9\% CC \numu interactions and 13.1\% CC \nue interactions. At the FD, assuming standard three-flavor oscillations, the efficiency of the sample is 87.6\% and the purity is 61.3\%, with the backgrounds comprising 73.8\% CC \numu interactions, 21.6\% CC \nue interactions, and 4.6\% CC \nutau interactions. A lower bound on the energy of the incident neutrino is estimated from the energy of the hadronic recoil system, with a mean resolution of 41.7\% on the energy of the recoil system in the FD.

We isolate a sample of CC \numu ($\numu{} N \rightarrow \mu X$) events by searching for interactions inside our detectors with a single outgoing $\mu$ track and possible hadronic activity from the recoil system $X$. We discriminate between CC and NC events by combining four topological variables describing track properties into a single discriminant variable, using a $k$-nearest-neighbour algorithm~\cite{ref:RustemThesis}. Events are required to have failed the NC selection procedure to be included in the CC \numu sample. In the ND, the selected CC sample has an efficiency of 53.9\% and a purity of 98.7\%, both estimated from a MC simulation. At the FD, assuming three-flavor oscillations, the corresponding efficiency is  84.6\% and the purity is 99.1\%. The neutrino energy is reconstructed by summing the energies of the muon and hadronic showers, with a mean resolution of 17.3\% in the FD.

\begin{figure}
\begin{center}
\includegraphics[width=\columnwidth]{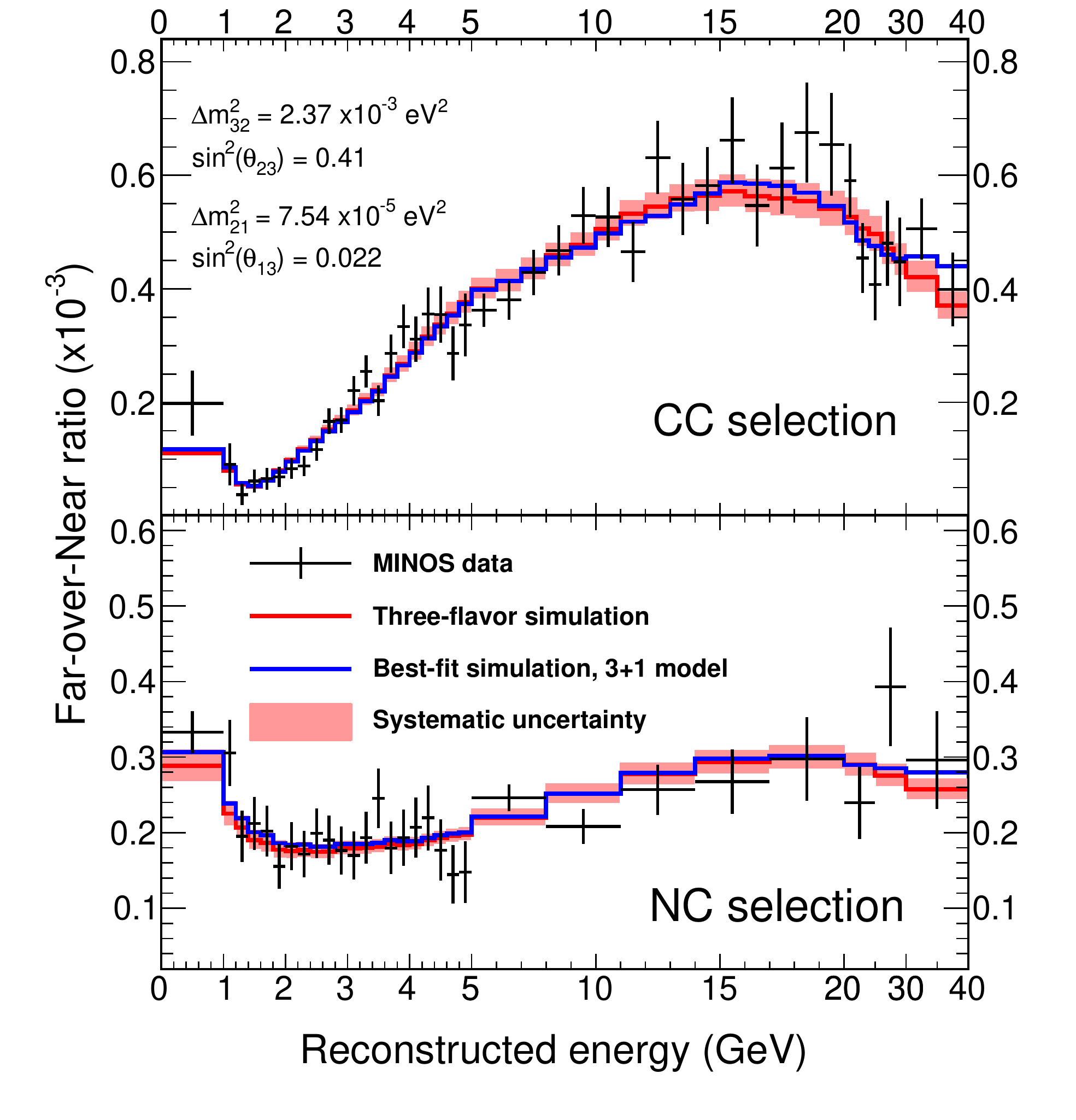}
\end{center}
\caption{\label{fig:fn} The ratios of the energy spectra in the far detector to those in the near detector, shown for the CC \numu (top) and NC (bottom) samples. The solid lines represent the predicted ratios from fits to the standard three-flavor oscillation model and to the 3+1 sterile neutrino model.}\end{figure}

MINOS oscillation analyses have traditionally used the CC and NC neutrino energy spectra measured by the ND to predict the spectra at the FD as a function of oscillation parameters~\cite{ref:MINOSCCPRD}.
However, the sterile oscillation parameter space to which MINOS is sensitive stretches over the range $10^{-3}\lesssim \dmsq{} \lesssim\unit[10^2]{eV^2}$, which could cause oscillations to impact both detectors~\cite{ref:Smirnov}. Therefore, instead of using the ND data to predict the FD energy spectra, we analyze the ratio of energy spectra observed in the FD to those observed in the ND. This FD-to-ND ratio is analyzed for both CC \numu and NC events, as shown in \mbox{Fig.\ \ref{fig:fn}}.
Aside from the overall difference in the number of events (caused by the distance between the detectors, their different masses and efficiencies, and the beam divergence) the main effect is the energy-dependent suppression of events at the FD caused by oscillations driven by $\Delta m^{2}_{32}$.
Our analysis searches for modulations on top of that oscillation pattern, caused by the sterile sector, by minimizing the $\chi^2$ as a function of the oscillation parameters:
\begin{equation} 
\chi^2_{CC,NC}   = \sum_{m=1}^{N}\sum_{n=1}^{N} (x_m - \mu_m) (V^{-1})_{mn} (x_n - \mu_n) + \mbox{const}. \label{eq:chi2}
\end{equation}
Here, we denote the measured FD-to-ND ratio as ${x_m}$, where $m=1,\ldots, N$ labels $N$ energy bins between 0 and \unit[40]{GeV}. The predicted ratio is denoted ${\mu_m}$. The dependence of $\mu_m$ on the oscillation parameters is taken from a MC simulation that includes the full range of experimental effects, and uses an exact form of all oscillation probabilities in vacuum with no approximations. In Eq.~(\ref{eq:chi2}), $V$ is an $N\times N$ covariance matrix expressing the combined statistical and systematic uncertainty on $\vec{\mu}$. For very high $\dmsq{41} \gtrsim \unit[50]{eV^2}$, both detectors are in the fast-oscillation limit and the only sensitivity comes from the overall rate measured in one of the two detectors. To account for the uncertainty on the overall rate we add a term $\chi^2_{\mathrm{rate}} = \frac{(X-M)^2}{\sigma^{2}_M}$, where $X$ and $M$ are the total number of ND events measured and simulated, respectively, and $\sigma_M$ is the uncertainty on $M$, which is conservatively assigned a value of 50\%, reflecting the fact that most measurements of neutrino fluxes and cross sections assume only three neutrino flavors.

We fit for $\theta_{23}$, $\theta_{24}$, $\theta_{34}$, $\dmsq{32}$ and $\dmsq{41}$, and hold all other parameters fixed. We set $\sin^2\theta_{12}=0.307$ and $\dmsq{21} = \unit[7.54\times10^{-5}]{eV^2}$ based on a global fit to neutrino data~\cite{ref:FogliGlobal}, and $\sin^2\theta_{13}=0.022$ based on a weighted average of recent results from reactor experiments~\cite{ref:DayaBay, RENO:2014jza, DC:2014dea}. An analysis of solar and reactor neutrino data yields the constraint $\sin^2\theta_{14} =|U_{e4}|^{2}<0.041$ at 90\% CL~\cite{ref:palazzoTh14}, which is small enough to have a negligible effect on this analysis, so we set $\theta_{14}=0$. This analysis has negligible sensitivity to $\delta_{13}$ and $\delta_{14}$, and minimal sensitivity to $\delta_{24}$; hence, all are set to zero. The impact of including the matter potential in the oscillation probability was investigated and found to have a negligible effect. The neutrino path length between the meson decay point and the ND was taken into account in the computation of oscillation probabilities.

\begin{figure}
\begin{center}
\includegraphics[width=\columnwidth]{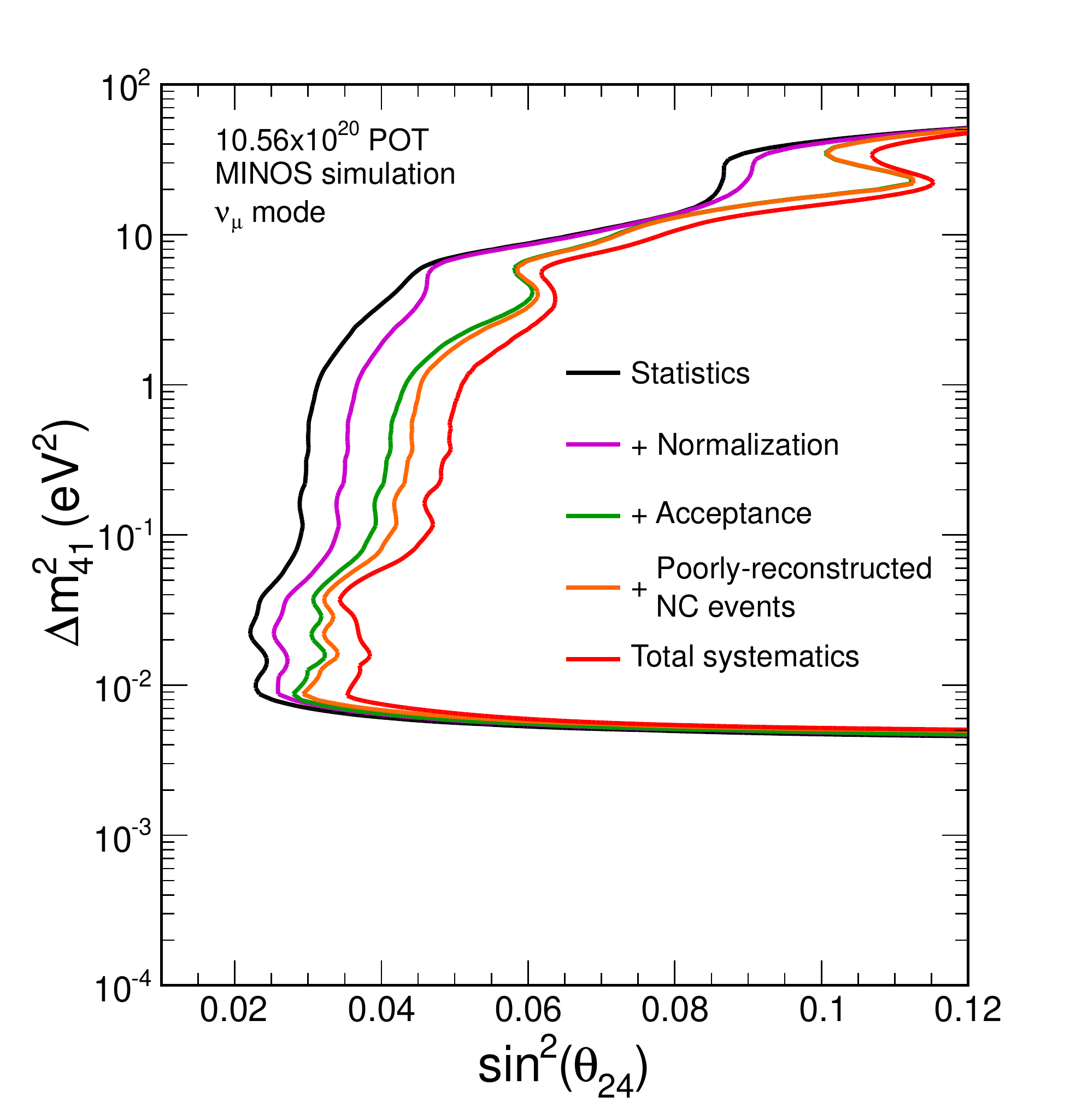}
\end{center}
\caption{\label{fig:syst} The effects of systematic uncertainties on the 90\% CL sensitivity in the $(\sin^{2}\theta_{24},\dmsq{41})$ plane, shown by successive inclusion of the listed uncertainties.} 
\end{figure}

Figure~\ref{fig:fn} shows a good agreement between the measured FD-to-ND ratios and those predicted using a three-flavor hypothesis.  No significant distortions indicative of sterile neutrinos are observed. The predicted ratios include both statistical and systematic uncertainties that are incorporated into Eq.~(\ref{eq:chi2}) via a covariance matrix,
\begin{equation}
\label{eq:CovarianceMatrix}
V = V_\mathrm{stat} + V_\mathrm{norm} +  V_\mathrm{acc} + V_\mathrm{NC} + V_\mathrm{other},
\end{equation}
where the terms account for the various sources of uncertainty.
Figure~\ref{fig:syst} shows the effects that the sources of systematic uncertainty have on the sensitivity of the sterile neutrino search. We describe each source of uncertainty below.

$V_{\mathrm{stat}}$ contains the statistical uncertainty, which is less than 24\% in each energy bin and 15\% on average. $V_{\mathrm{norm}}$ contains a 1.6\% uncertainty in the relative normalization of the CC sample between the ND and FD, and a corresponding 2.2\% uncertainty for the NC sample. This accounts for uncertainties in reconstruction efficiencies. It was determined by a study in which a team of scanners looked at events in both detectors from both simulation and data to assess the level of reconstruction failures. No evidence for a mismodeling of the reconstruction failures was observed, and the values quoted for the uncertainties are the statistical precision to which the modeling could be tested.

$V_{\mathrm{acc}}$ accounts for uncertainties on the acceptance and selection efficiency of the ND. These uncertainties were evaluated by varying event-selection requirements in the data and MC simulation to probe known weaknesses in the simulation. As these requirements were varied, the total variations in the ND data to MC ratios were taken as systematic uncertainties on the FD-to-ND ratios. The total uncertainty included in $V_{\mathrm{acc}}$, which is energy dependent and includes correlations between different bins, varies from $2\%$ to $6\%$ for the CC sample and is below 0.6\% at all energies for the NC sample.

$V_{\mathrm{NC}}$ accounts for an uncertainty on the procedure used to remove poorly reconstructed events from the NC sample. The variables used to identify such poorly reconstructed events are not perfectly modeled by the MC simulation. A procedure, described in Ref.~\cite{ref:AlenaThesis}, assesses an uncertainty arising from this mismodeling. The total uncertainty, which includes correlations between energy bins, falls from $5\%$ below \unit[1]{GeV} to less than $1.5\%$ above $\unit[5]{GeV}$.

$V_{\mathrm{other}}$ includes terms to account for all sources of uncertainty in neutrino interaction cross sections and the flux of neutrinos produced in the NuMI beam. The total uncertainty on the FD-to-ND ratios arising from these sources is no more than $4\%$ in any parts of the energy spectra.

\begin{figure}
\begin{center}
\includegraphics[width=\columnwidth]{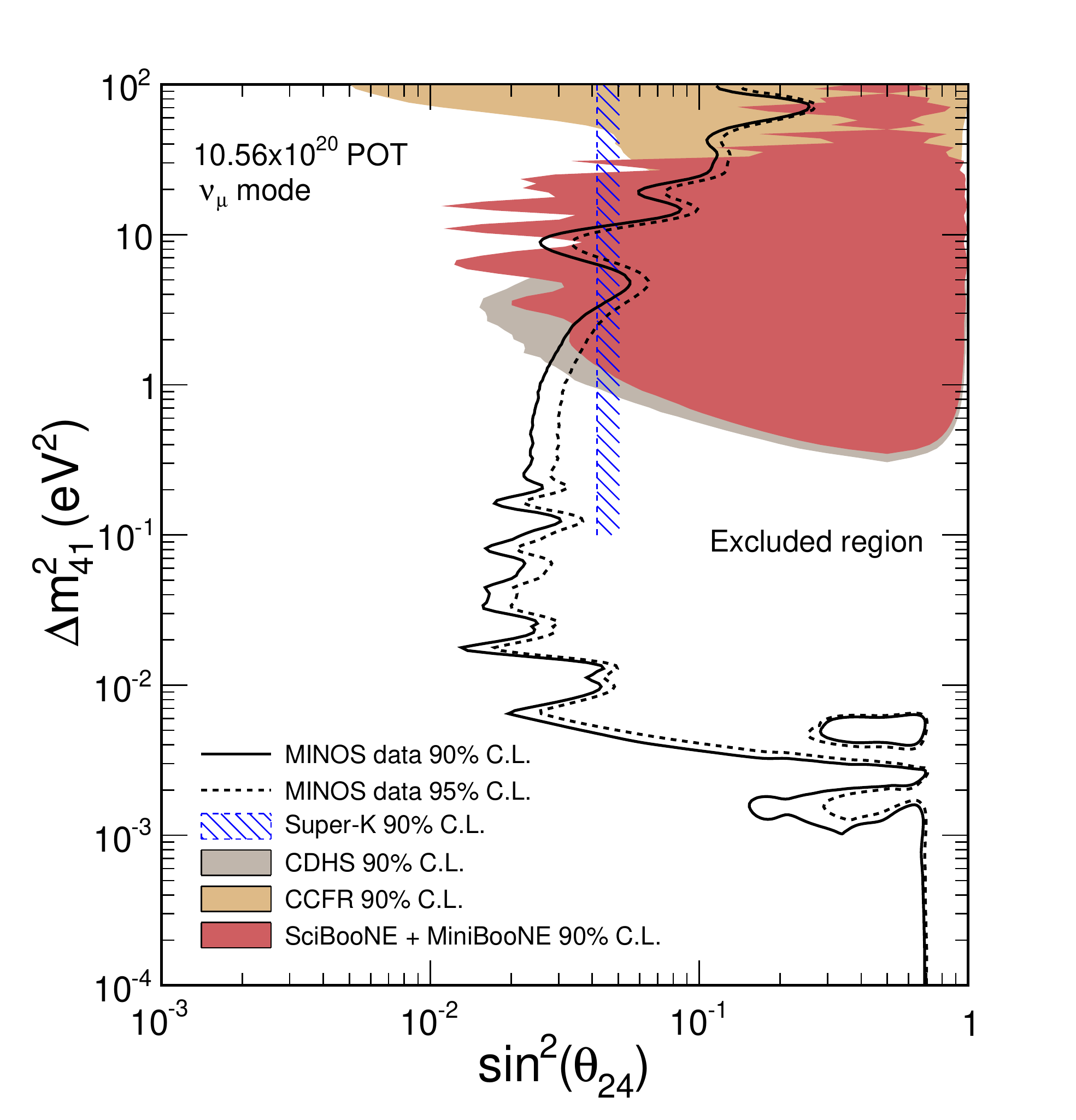}
\end{center}
\caption{\label{fig:result} The MINOS 90\% and 95\% confidence limits in the $(\sin^2\theta_{24},\dmsq{41})$ plane compared with results from previous experiments~\cite{ref:SuperK,ref:CDHS,ref:CCFR,ref:SciBooNE}. The areas to the right of the MINOS lines are excluded at their respective confidence levels.}
\end{figure}

We fit the $3+1$ model to the data by dividing the $(\sin^2\theta_{24},\ \dmsq{41})$ plane into fine bins and minimizing Eq.~(\ref{eq:chi2}) in each bin with respect to \dmsq{32}, $\theta_{23}$, and $\theta_{34}$. 
At each point in the plane we interpret the significance of the $\Delta\chi^2$ with respect to the global minimum according to the unified procedure of Feldman and Cousins~\cite{ref:FC}. In this procedure, MC pseudoexperiments are generated, with bin-to-bin statistical and systematic fluctuations incorporated by sampling from a multidimensional Gaussian with covariance matrix $V$ [defined in Eq.~(\ref{eq:CovarianceMatrix})].  The result is shown in \mbox{Fig.\ \ref{fig:result}}, with the area to the right of the curves excluded at their respective confidence limits. The data are consistent with three-flavor oscillations at 54.7\% CL; no evidence for sterile neutrinos is observed. The world's best limit on $\sin^2\theta_{24}$ is established for $\dmsq{41}<\unit[1]{eV^2}$, a largely unmeasured region of parameter space.

The limit obtained from the data is stronger than expected from the sensitivity, as can be seen from a comparison of Figs.~\ref{fig:syst} and~\ref{fig:result}. A study shows that 8\% of fake experiments obtain an exclusion stronger than that obtained from the data at $\Delta m^{2}_{41}=\unit[1]{eV^{2}}$. At $\Delta m^{2}_{41} = \unit[0.5]{eV^{2}}$, the CC sample provides $75\%$ of the $\Delta \chi^{2}$ that gives rise to the \unit[90\%]{C.L.} exclusion contour, with the NC sample providing the remaining $25\%$.

For $\dmsq{41}<\unit[10^{-2}]{eV^2}$ it becomes possible for one of the three mass splittings \dmsq{41}, \dmsq{42} or \dmsq{43} to match the scale of oscillations in the $\Delta m^{2}_{32}$ sector. This results in solutions that are degenerate with the standard three-flavor prediction, creating an island of allowed parameter space that is visible in \mbox{Fig.\ \ref{fig:result}}.

Upper limits on the angles $\theta_{24}$ and $\theta_{34}$, which correspond to limits on elements of the PMNS matrix, may be defined at fixed values of \dmsq{41}. For $\dmsq{41}=\unit[0.5]{eV^{2}}$, the data constrain $\sin^2\theta_{24}< [0.016\,\mathrm{(\unit[90\%]{CL})}, 0.022\,\mathrm{(\unit[95\%]{CL})}]$; under the assumption that $|U_{e4}|^{2}=0$, these are also limits on $|U_{\mu 4}|^{2}=c^{2}_{14}s^{2}_{24}$. For $\dmsq{41}=\unit[0.5]{eV^{2}}$, the data also constrain $\sin^2\theta_{34}<[0.20\,\mathrm{(\unit[90\%]{CL})}, 0.28\,\mathrm{(\unit[95\%]{CL})}]$; under the assumption $c^{2}_{14}=c^{2}_{24}=1$, these are also limits on $|U_{\tau 4}|^{2}=c^{2}_{14}c^{2}_{24}s^{2}_{34}$.

In conclusion, we have used samples of CC \numu and NC interactions from the NuMI neutrino beam to place a constraint on the existence of sterile neutrinos. We use a $3+1$ model to quantify this constraint, and are sensitive to a range of $\Delta m^{2}_{41}$ covering almost 5 orders in magnitude. Over much of this region, we place the first constraints on the mixing angle $\theta_{24}$. In an accompanying Letter~\cite{ref:DayaBayCombination}, we present a combination of this constraint with those on $\theta_{14}$ from the Daya Bay~\cite{ref:DayaBay8AD} and Bugey~\cite{ref:Bugey} reactor experiments to set a limit that is directly comparable with the possible hints of sterile neutrinos seen by the LSND and MiniBooNE experiments.

This work was supported by the U.S. DOE, the U.K. STFC, the U.S. NSF, the State and University of Minnesota, and Brazil's FAPESP, CNPq and CAPES.  We are grateful to the Minnesota Department of Natural Resources and the personnel of the Soudan Laboratory and Fermilab. We thank the Texas Advanced Computing Center at The University of Texas at Austin for the provision of computing resources. We thank A.~Smirnov and C.~Giunti for useful discussions.\newline

\emph{Note Added}.---A paper by the IceCube Collaboration that sets limits using sterile-driven disappearance of muon neutrinos has recently appeared~\cite{ref:IceCube}. The results place strong constraints on $\sin^2 2\theta_{24}$ for $\Delta m^2_{41} \in (0.1,10)\ {\rm eV}^2$. Furthermore, a paper that reanalyses the same IceCube data in a model including nonstandard neutrino interactions has also recently appeared~\cite{ref:IceCubeNSI}.

\bibliography{paper}


\end{document}